\documentclass[aps, prl, amsfonts, amsmath, amssymb, reprint, nofootinbib, twoside, superscriptaddress, letterpaper, floatfix]{revtex4-2}

\usepackage{docmute}   
\usepackage{graphicx}
\usepackage{dcolumn}
\usepackage{bm}
\usepackage[version=4]{mhchem}
\usepackage{amsmath,amssymb,amsfonts}
\usepackage{siunitx}
\usepackage{placeins}
\DeclareSIUnit{\wv}{\%\,w/v}
\DeclareSIUnit{\square}{sq}
\DeclareSIUnit{\pix}{pixel}
\DeclareSIUnit{\fps}{fps}
\DeclareSIUnit{\radians}{rad}
\DeclareSIUnit{\elecVolt}{eV}
\DeclareSIUnit{\molar}{M}
\DeclareSIUnit{\bar}{bar}
\usepackage[dvipsnames]{xcolor}
\usepackage{utfsym}

\usepackage{comment}
\newcommand{\ag}[1]{}
\newcommand{\mc}[1]{}
\newcommand{\dl}[1]{}
\newcommand{\ut}[1]{}
\newcommand{\li}[1]{}

\usepackage{hyperref}
\hypersetup{colorlinks=true, linkcolor=blue, citecolor=blue, filecolor=magenta, urlcolor=blue}

\begin{document}


\raggedbottom
\title{Delay-induced current reversal in traveling activity waves}

\author{Ueli T\"opfer}
    \affiliation{Department of Materials, ETH Zurich, 8049 Zurich, Switzerland}

\author{Adri\`a Garc\'es}
    \affiliation{Departament de F\'\i sica de la Mat\`eria Condensada, Universitat de Barcelona, Mart\'\i{} i Franqu\`es 1, E08028 Barcelona, Spain}
    \affiliation{University of Barcelona Institute of Complex Systems (UBICS), Mart\'\i{} i Franqu\`es 1, E08028 Barcelona, Spain}
    \affiliation{Computing and Understanding Collective Action (CUCA) Lab, Mart\'\i{} i Franqu\`es 1, E08028 Barcelona, Spain}

\author{Miguel Chamorro-Burgos}
    \affiliation{F\'\i sica Te\'orica, Universidad de Sevilla, Apdo.~1065, 41080 Sevilla, Spain}

\author{Demian Levis}
    \affiliation{Departament de F\'\i sica de la Mat\`eria Condensada, Universitat de Barcelona, Mart\'\i{} i Franqu\`es 1, E08028 Barcelona, Spain}
    \affiliation{University of Barcelona Institute of Complex Systems (UBICS), Mart\'\i{} i Franqu\`es 1, E08028 Barcelona, Spain}
    \affiliation{Computing and Understanding Collective Action (CUCA) Lab, Mart\'\i{} i Franqu\`es 1, E08028 Barcelona, Spain}

\author{Alvaro Dom\'\i nguez}
    \affiliation{F\'\i sica Te\'orica, Universidad de Sevilla, Apdo.~1065, 41080 Sevilla, Spain}
    \affiliation{Instituto Carlos I de F\'\i sica Te\'orica y Computacional, 18071 Granada, Spain}

\author{Ignacio Pagonabarraga}
    \affiliation{Departament de F\'\i sica de la Mat\`eria Condensada, Universitat de Barcelona, Mart\'\i{} i Franqu\`es 1, E08028 Barcelona, Spain}
    \affiliation{University of Barcelona Institute of Complex Systems (UBICS), Mart\'\i{} i Franqu\`es 1, E08028 Barcelona, Spain}

\author{Lucio Isa}
    \email{lucio.isa@mat.ethz.ch}
    \affiliation{Department of Materials, ETH Zurich, 8049 Zurich, Switzerland}

\begin{abstract}
Self-propelled particles that regulate their speed based on a response to a local activity source can move with or against a traveling wave of that source depending on their response time. Silica-titania Janus microswimmers, whose UV-controlled speed relaxes with a delay intrinsic to the titania photoconductivity adaptation, carry two independent memory channels: the familiar directional memory dictated by persistent active Brownian motion, and a distinct memory from the speed adaptation. Combining experiments, Langevin simulations, and a moment expansion of the corresponding Fokker-Planck equation, we show that this delayed response to speed adaptation reshapes transport under traveling UV-illumination waves and can even reverse it relative to the direction set by orientational memory alone. In particular, we observe that delay promotes transport in the direction opposite to the traveling activity waves, where, beyond a critical value, the particle current reverses with respect to the wave propagation direction. Crucially, this behavior requires an asymmetric response to the speed adaptation, with an instantaneous rise but a delayed fall upon exposure to time-varying UV light. Theory further shows that this phenomenology is specific to the response asymmetry rather than to the mere presence of a delay, i.e. reversing the asymmetry reverses the effect of the delay on the particle current relative to the wave. We thus establish  propulsion-delay asymmetry as a control parameter for directed transport of responsive active particles in traveling activity landscapes.
\end{abstract}

\maketitle

A finite response time is central to how bacteria sense and navigate chemical gradients in their environment. While the concentration difference across a single cell's own length is small, a cell can probe a far larger stretch of the gradient than its body alone would allow by comparing concentrations sensed at different times as it swims~\cite{Segall1986, deGennes2004}.

More generally, when a microswimmer, biological or synthetic, adjusts its motion in response to a time-varying activity cue, such as a traveling chemical wave, a finite response time can give rise to net transport up or downstream the traveling pattern. The direction and magnitude of this transport depend on a competition between timescales: on one side, the wave period $\tau_\ell\equiv\ell/c$, set by the wavelength $\ell$ and the wave speed $c$, and on the other side, the timescales controlling the microswimmer motion, namely the swim time $\tau_u\equiv\ell/u_{\max}$ needed to cross one wavelength at its maximum speed $u_{\max}$, and the internal memory times over which the microswimmer retains information about its past, here its orientational persistence and the delay in adapting its propulsion.

For instance, in the chemotactic amoeba \textit{Dictyostelium discoideum}, the relevant memory is the decay time of the chemotactic response. When it matches the wave period, cells are desensitized to the trailing edge of a self-generated, outward-traveling chemoattractant wave~\cite{Tomchik1981}, which rectifies their motion toward the source and allows them to aggregate~\cite{Hfer1994, Goldstein1996}.

For active Brownian particles whose speed adapts instantaneously, an orientational memory is anyway present and determined by rotational diffusion: the particles self-propel with persistent motion with a rotational decorrelation time $\tau_r\equiv1/D_r$, where $D_r$ is the rotational diffusion coefficient. With no deterministic coupling to any local gradient, orientational memory alone drives transport against slow waves and along fast ones, with $\tau_r$ setting the wave speed at which the effect peaks~\cite{Maggi2018, Geiseler2016, Geiseler2017a, Geiseler2017b}. Alternatively, the presence of a deterministic aligning torque can lead to the same transport reversal by a different route: here, memory emerges in the form of alignment time, which competes with the time spent inside a moving activity pulse, as demonstrated for Janus colloids whose orientation aligns toward a light gradient~\cite{Lozano2019}.

In all these cases, the memory resides in the orientation or, for \textit{Dictyostelium}, in a chemotactic response. A self-propelled particle can, however, carry a second, independent memory: a delay $\tau$ in the adaptation of its own propulsion speed in response to a local variation of activity. In most synthetic swimmers with adaptive propulsion this delay is small and typically neglected~\cite{Palacci2014,Zehavi2022}. In some systems, however, it is large enough to matter, as for genetically engineered photokinetic bacteria~\cite{Frangipane2018,Arlt2018} and for photoconductive Janus particles, where speed adaptation to light modulation is an intrinsic material response~\cite{Tpfer2025}. The two memories differ in kind. Orientational persistence locks in the particle's own past direction up to $\tau_r$ and is blind to an activity wave, whereas a propulsion delay records the wave's recent state, so that propulsion reflects where the wave was rather than where the particle  is~\cite{Loos2019}. Imposed delays are known to induce and reverse net currents in passive Brownian ratchets~\cite{Feito2008, Loos2014} and have been engineered into active agents through feedback loops~\cite{Mijalkov2016, Leyman2018,FernandezRodriguez2020, MassanaCid2022, Holubec2026, HolubecReview2026}. Delayed speed adaptation, including exponential relaxation, was further predicted to rectify transport in spatially asymmetric, periodic activity landscapes and to reverse it upon tuning the delay~\cite{Rein2026}. All these studies, however, consider static or feedback-generated activity landscapes~\cite{Holubec2026, MassanaCid2022, Rein2026}, not an externally imposed traveling wave, which the particle intrinsically responds to.

In this Letter, combining experiments, Langevin simulations, and a moment expansion of the Fokker-Planck equation, we show that delay biases transport against the wave and that, beyond a critical value, reverses the particle current. We also show that current reversal requires an asymmetric time response.

\paragraph{Experimental system and simulations.} We use silica-titania Janus microswimmers, \SI{2}{\um} in diameter, suspended in water with Pluronic F-127 and sedimented onto the lower of two ITO electrodes confining a thin liquid cell. A uniform AC electric field propels them by induced-charge electrophoresis~\cite{Boymelgreen2022, Squires2006}, while the photoconductive titania cap lets UV illumination modulate the propulsion speed orthogonally to the driving field~\cite{Tpfer2025}. Projecting UV patterns into the cell with a digital micromirror device (DMD) thus lets us impose spatiotemporal activity landscapes on the swimmers. 

In these systems the propulsion speed $v$ grows with light intensity up to a saturation level, Fig.~\ref{fig:setup}(a), as previously reported for photocatalytic microswimmers~\cite{Palacci2013,Chen2016,Aubret2018}. Propulsion does not, however, respond instantaneously to the removal of illumination. It relaxes toward its target value with an intrinsic delay set by the photoconductivity of the titania and the lifetime of its charge carriers~\cite{Tpfer2025, Litke2017a, Litke2017b}. Switching illumination on and off directly reveals this memory in speed: switching the UV on, the active speed rises essentially instantaneously (see SM \cite{SM}, Fig.~S1); switching it off, it relaxes smoothly, following a single exponential decay with a time constant $\tau\approx\SI{2}{\second}$, Fig.~\ref{fig:setup}(b). This asymmetric response time adds on to a timescale characteristic of the decorrelation of the particle orientation set by Brownian rotational diffusion, $\tau_r\approx\SI{6.3}{\second}$. 
The DMD moves the illumination pattern as a traveling wave, Fig.~\ref{fig:setup}(c,d). As the particles are illuminated, they adjust their propulsion speed to a local ``target speed'' set by the local light intensity and retain it over a time $\sim\tau$ once the intensity decreases.

\begin{figure}[ht]
\centering
\includegraphics[width=0.99\columnwidth]{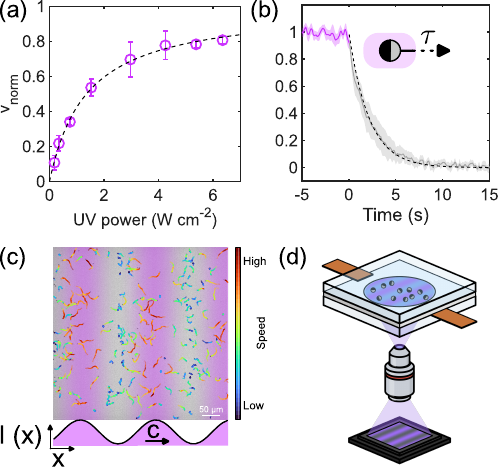}
\caption{\textbf{Propulsion response of the Janus microswimmers to UV illumination.}
(\textbf{a}) Active speed vs. light intensity, normalized by the maximum speed, from MSD fits across a range of illumination intensities; error bars are standard deviations. Dashed line: fit to a saturating hyperbolic response. Data redrawn from Ref.~\cite{Tpfer2025}.
(\textbf{b}) Active speed following a spatially uniform switch-off of UV illumination, normalized and averaged over three illumination cycles across four experimental cells; shaded regions are 95\% confidence intervals. Dashed line: fitted exponential decay, giving a delay of \SI{2}{\second}.
(\textbf{c}) Micrograph of microswimmers under the traveling-wave pattern at steady state ($u_{\max}=\SI{8}{\um\per\second}$, $c=\SI{0.6}{\um\per\second}$), showing particle trajectory tails over the preceding \SI{5}{\second}, color-coded by instantaneous speed, with a sketch below showing the wave intensity profile.
(\textbf{d}) Schematic of the setup: the particle suspension is confined between two ITO-coated glass slides, contacted with copper tape to apply the AC field, and a DMD projects the light pattern through the objective into the sample plane.}
\label{fig:setup}
\end{figure}

Correspondingly, we model the system as overdamped active Brownian particles with the same asymmetric, first-order propulsion lag, driven by a sinusoidal traveling wave (simulation details in the Supplemental Material~\cite{SM}).

\begin{figure*}[t]
\centering
\includegraphics[width=0.99\textwidth]{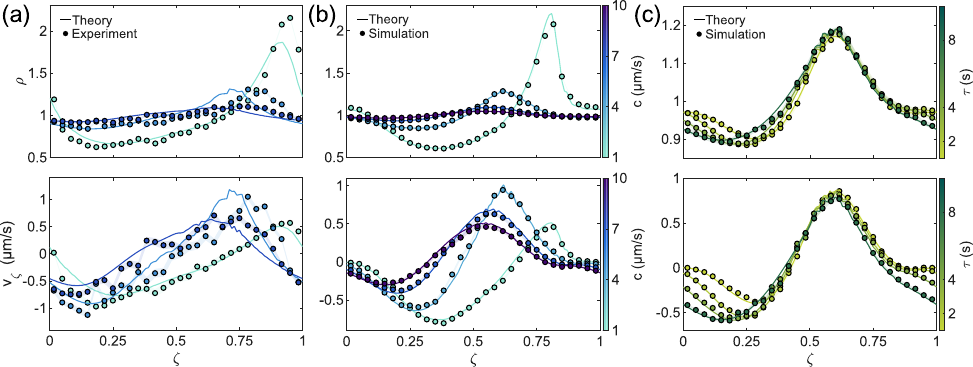}
\caption{\textbf{Density and local mean velocity: experiment, simulation, and theory.}
Top row, $\rho(\zeta)$; bottom row, $v_\zeta$. Markers: experimental (a) or simulated (b, c) data; solid lines: moment expansion.
(\textbf{a}) Experiments and theory for different wave speed $c$ (color bar), for $u_{\max}$ between \SIrange{6}{8}{\um\per\second}.
(\textbf{b}) Simulation at fixed $u_{\max}=\SI{5}{\um\per\second}$, $\tau=\SI{2}{\second}$, for different $c$ (color bar).
(\textbf{c}) Simulation at fixed $u_{\max}=\SI{5}{\um\per\second}$, $c=\SI{5}{\um\per\second}$, for different delay $\tau$ (color bar).}
\label{fig:pde_sim_exp}
\end{figure*}

\paragraph{Model.} In the frame comoving with the wave, $z\equiv x-ct$, with $x$ the lab-frame position, the particle's Langevin dynamics for orientation $\theta$ and active speed $v$ reads
\begin{align}
\dot z &= v\cos\theta - c + \sqrt{2D}\,\eta, \notag\\
\dot\theta &= \sqrt{2D_r}\,\chi, \notag\\
\dot v &= -\tau(z,v)^{-1}\big[v-u(z)\big],
\end{align}
with $\eta,\chi$ independent Gaussian white noises, $D,D_r$ the translational and rotational diffusion coefficients, $u(z)$ the traveling pattern's target speed, and $\tau(z,v)$ the response time measured above: an instantaneous rise ($\tau_{\mathrm{on}}\approx0$) and a delayed fall ($\tau_{\mathrm{off}}\equiv\tau$). When $\tau=0$, $v=u(z)$ exactly and the standard active Brownian particle model is recovered. A finite $\tau$ is what lets $v$ carry memory of the particle's own history. The Fokker-Planck equation for the joint probability density $\psi(z,\theta,v,t)\equiv\langle\delta(z-z_t)\delta(\theta-\theta_t)\delta(v-v_t)\rangle$ follows directly from these Langevin equations:
\begin{align}
\partial_t\psi = &-\partial_z\big[(v\cos\theta-c)\psi\big] + D\partial_z^2\psi + D_r\partial_\theta^2\psi \notag\\
&+ \partial_v\left[\left(\frac{v-u(z)}{\tau(z,v)}\right)\psi\right]. \label{eq:fp_z}
\end{align}
The first three terms in Eq.~\eqref{eq:fp_z} are advection along $z$ at the particle's own speed relative to the wave, translational diffusion, and rotational diffusion of the orientation; the fourth is the relaxation of the active speed toward the local wave target just introduced above. We seek steady-state solutions of the equation that preserve the symmetry $\theta\to-\theta$, which allows us to expand $\psi$ in cosine harmonics, $\Psi_k\equiv\int \cos(k\theta)\,\psi\,d\theta$. This reduces the problem to a hierarchy of moment equations, which we close at second order ($\Psi_3\equiv0$), retaining the particle number density ($\Psi_0$), the polar order ($\Psi_1$), and the nematic order ($\Psi_2$). Its root mean square amplitude grows with increasing persistence length in the simulations (Fig.~S3), and the second-order closure reproduces this trend with good agreement, consistent with general studies of active Brownian particle moment closures~\cite{Gautry2026} (the full scaling argument is given in the Supplemental Material~\cite{SM}). We solve the resulting closed linear system numerically as a sparse linear system without time-stepping (full scheme in the Supplemental Material~\cite{SM}), and the solution is then nondimensionalized in space for comparison with experiments and simulations, with $\zeta\equiv z/\ell$ the resulting dimensionless, periodic phase. This gives us direct access to the phase-resolved density $\rho(\zeta)\equiv\ell\int\Psi_0\,dv ,$ and, from the polar-order flux $q(\zeta)\equiv\ell\int v\,\Psi_1\,dv$, the local mean velocity $v_\zeta=q(\zeta)/\rho(\zeta)$ along the wave at phase $\zeta$, without free parameters beyond those fixed by the experiments or simulations.

\paragraph{Results.} Figure~\ref{fig:pde_sim_exp} shows the phase-resolved density and local mean velocity from experiments (panel a, markers), simulations (panels b, c, markers), and theory (solid lines). For the experimental panels, the theory uses the illumination profile $u(\zeta)$ measured directly from the data, rather than the idealized sinusoid prescribed by the DMD. This measured profile automatically captures the particles' actual speed response, including the nonlinear relation between light intensity and active speed; its derivation, together with how the instantaneous velocity profiles are debiased for the influence of diffusion using a Rice distribution, is given in the Supplemental Material~\cite{SM}. The theory reproduces both quantities closely across the full range of tested wave speeds and delays. Both experiments and simulations fall also close to the predicted peak and trough values in parity plots given in the Supplemental Material (Fig.~S4)~\cite{SM}.

At low wave speed, density peaks just before $\zeta=1$, where the driving profile itself vanishes ($u(\zeta)\to0$), as particles slow toward this point and linger. This is the classic result that active particles in a spatially-varying speed field accumulate in slow regions, with density $\rho(\mathbf{r})\propto1/v(\mathbf{r})$~\cite{Schnitzer1993,Tailleur2008}, exploited directly for density shaping in photokinetic bacteria~\cite{Frangipane2018} and in our own silica-titania microswimmers~\cite{Tpfer2025}. This result is also recovered analytically as the low-wave-speed limit of the same traveling-activity-wave problem~\cite{Maggi2018}. As wave speed grows and approaches the particles' own speed, this quasi-static picture breaks down and the peak weakens, lowering and shifting away from $\zeta=1$ while the trough shallows and shifts further behind. This is the same crossover toward a flattened, near-uniform density predicted by the same theory at high wave speed~\cite{Maggi2018}. The density peak has little sensitivity to delay at fixed wave and particle speed, since its position is set almost entirely by how much the particles' own speed exceeds the wave speed. This follows directly from the asymmetric relaxation itself. The peak forms as particles enter the profile under the fixed, near-instantaneous $\tau_{\mathrm{on}}$, so it changes little as $\tau_{\mathrm{off}}=\tau$ is swept, while only the trailing edge, governed by $\tau_{\mathrm{off}}$, broadens as the delay grows.

The local mean velocity $v_\zeta$ shows the same two-sided structure in both experiment and simulation, with a positive peak just ahead of the wave crest, driven forward by the wave, and a negative trough behind it, where particles fall back. As wave speed increases, the peak first rises, then falls again while shifting toward the crest, and the trough shallows while shifting further behind, mirroring the density profile's own evolution. Delay leaves the peak essentially unchanged, consistent with the same $\tau_{\mathrm{on}}$-governed insensitivity seen in the density, but deepens the trough at first, saturating before it broadens further with increasing $\tau$, on the $\tau_{\mathrm{off}}$-governed side of the profile.

\begin{figure}[ht]
\centering
\includegraphics[width=0.9\columnwidth]{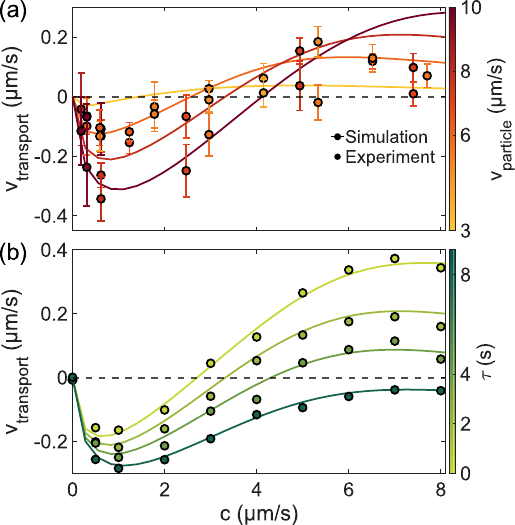}
\caption{\textbf{Net drift velocity $v_{\mathrm{transport}}$ vs. wave speed $c$.} Markers: experimental (a) or simulated (b) data; solid line: moment expansion.
(\textbf{a}) Experimental $v_{\mathrm{transport}}$ vs. $c$, colored by particle speed $u_{\max}$, at fixed $\tau=\SI{2}{\second}$.
(\textbf{b}) Simulated $v_{\mathrm{transport}}$ vs. $c$, colored by delay $\tau$, at fixed $u_{\max}=\SI{8}{\um\per\second}$.}
\label{fig:transport}
\end{figure}

The quantity of central interest is the net drift velocity, i.e., the net displacement per unit time that the traveling wave imparts on the particles, obtained as the density-weighted average of $v_\zeta$ over one period,
\begin{equation}
v_{\mathrm{transport}} = \int_0^1 \rho(\zeta)\,v_\zeta\,d\zeta.
\end{equation}
We take $v_{\mathrm{transport}}>0$ for a net particle current along the direction of wave propagation and $v_{\mathrm{transport}}<0$ for net transport against it. Equivalently, $v_{\mathrm{transport}}=\ell J$, with $J$ the stationary probability current~\cite{Maggi2018}, so that a sign change of $v_{\mathrm{transport}}$ constitutes a current reversal.

\begin{figure*}[t]
\centering
\includegraphics[width=0.99\textwidth]{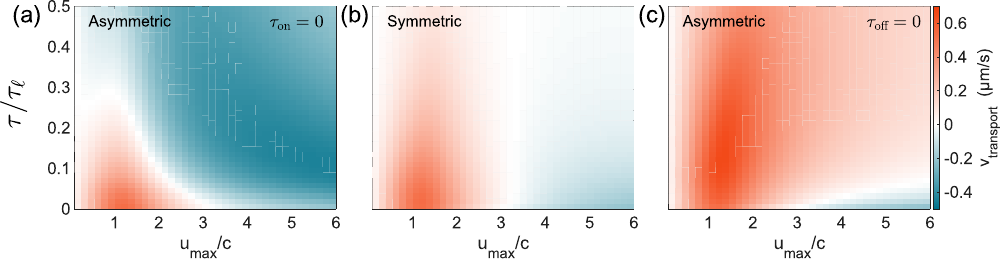}
\caption{\textbf{Net drift velocity for asymmetric, symmetric, and reversed-asymmetric delay.}
Theoretical net drift velocity $v_{\mathrm{transport}}$ (color, \si{\um\per\second}, color map centered at zero), solved directly from the moment expansion of Eq.~\eqref{eq:fp_z} at fixed particle radius $R=\SI{1.03}{\um}$ (giving $\tau_D/\tau_r\approx5.66\times10^4$) and wavelength $\ell=\SI{200}{\um}$.
(\textbf{a}) Asymmetric delay ($\tau_{\mathrm{on}}\approx0$, $\tau_{\mathrm{off}}=\tau$), as used throughout this work; reproduces the same regime as Fig.~\ref{fig:transport}.
(\textbf{b}) Symmetric delay ($\tau_{\mathrm{on}}=\tau_{\mathrm{off}}=\tau$).
(\textbf{c}) Reversed asymmetric delay ($\tau_{\mathrm{on}}=\tau$, $\tau_{\mathrm{off}}\approx0$).}
\label{fig:sym_asym}
\end{figure*}

Figure~\ref{fig:transport}(a) shows $v_{\mathrm{transport}}$ vs. wave speed. Experimental data, overlaid with the theory at fixed $\tau=\SI{2}{\second}$, agree qualitatively, with the direction of net transport reversing with wave speed, running against a slow wave and along with a fast wave, respectively. This is the same crossover already established for orientational memory alone, without any delay in speed~\cite{Maggi2018,Geiseler2016,Geiseler2017a,Geiseler2017b}. The role of the delay becomes clear once $\tau$ is varied at fixed particle speed in the simulations, Fig.~\ref{fig:transport}(b). The curve's overall shape, and the wave speeds at which it reaches its maximum and minimum, change little with $\tau$. Instead, increasing $\tau$ shifts the whole curve downward, more strongly at its maximum than at its minimum. The shift continues until, for sufficiently large $\tau$, transport with the wave is eliminated entirely, leading to a net current against the wave across the full range tested. This behavior follows directly from the phase-resolved behavior established above: delay leaves the leading part of the $\rho$ and $v_\zeta$ profiles largely unaffected while deepening their trailing part, shifting the density-weighted net transport against the wave as $\tau$ grows.
 
Besides the timescales introduced above, there is also the translational-diffusion time across one wavelength, $\tau_D\equiv\ell^2/D$. For a given profile shape $u(\zeta)/u_{\max}$, the transport is thus fully set by $u_{\max}/c=\tau_\ell/\tau_u$ and the ratios $\tau/\tau_\ell$, $\tau_D/\tau_\ell$, and $\tau_D/\tau_r$. The latter ratio depends only on particle size and wavelength, which are fixed in our simulations and experiments. Fig.~\ref{fig:sym_asym} therefore maps the net transport across the $(u_{\max}/c)$-$(\tau/\tau_\ell)$ plane, beyond the wave speeds and delays tested experimentally, isolating the role of the delay's asymmetry. Here $\tau/\tau_\ell$ measures how large a fraction of one wave-period the delay spans. Meanwhile, $\tau_D/\tau_\ell>10^3$ across the entire range, so translational diffusion never competes with the wave-crossing timescale. Our experimental setup itself restricts $\tau/\tau_\ell$ to below $\approx0.1$. The used wavelength $\ell=\SI{200}{\um}$ balances the DMD dithering resolution against fitting several wavelengths within the field of view, and $c$ is limited by the DMD's $\approx\SI{50}{\milli\second}$ update time needed to move the activity wave smoothly. $u_{\max}/c$, by contrast, reaches up to $\approx50$ in the experiments.
 
With the asymmetric delay used throughout this work, Fig.~\ref{fig:sym_asym}(a), transport with the wave survives only in a narrow region around $u_{\max}/c\approx1$ and weakens as $u_{\max}/c$ moves away from this value in either direction, or as $\tau/\tau_\ell$ grows. Across most of the plane, particles are transported against the wave, a regime reached by increasing either $u_{\max}/c$ or $\tau/\tau_\ell$, so delay alone can reverse the particle flow with respect to the wave propagation direction. 
Making the delay symmetric instead, Fig.~\ref{fig:sym_asym}(b), removes this ability entirely: increasing $\tau/\tau_\ell$ only ever weakens transport in whichever direction is already set by $u_{\max}/c$, so delay becomes uniformly detrimental to transport magnitude, and reversal is possible only by varying $u_{\max}/c$, never the delay itself. Conversely, reversing the asymmetry, Fig.~\ref{fig:sym_asym}(c), flips the picture entirely: the system is now dominated by transport with the wave, which is enhanced by delay rather than suppressed, with transport against the wave confined to a small region at $u_{\max}/c\gtrsim3$ and very small delays. Together, the three panels show that delay does not affect transport indiscriminately. It becomes a tuning parameter whose effect depends on the details of its functional form, including its sign and symmetry.

\paragraph{Conclusion.} We have shown that silica-titania Janus microswimmers carry an additional distinct memory channel in speed adaptation beyond persistent active Brownian motion. This delay strongly reshapes the transport set by orientational memory alone in a traveling activity wave and can even reverse the particle flow with respect to the wave propagation direction. Analysis of the form of speed relaxation shows that the asymmetry between velocity rise and fall, not delay alone, is the genuine ingredient rectifying transport, in analogy to ratchets driven by temporally asymmetric zero-mean noise~\cite{Luczka1995}. Reversing this asymmetry reverses its effect on transport entirely. This establishes the asymmetry of propulsion delay as a control parameter for directed transport that acts independently of the wave itself, and suggests that any system with an intrinsically delayed response to a traveling stimulus, biological or synthetic, may exploit the same competition between wave speed, delay, and persistence time.

\begin{acknowledgments}
U.T. thanks K.~Studer, N.~Windbacher, and R.~Style for their time and for many good discussions. D.L. acknowledges A.~Katona, S.~L\'evay, and I.~Zurigel for insightful discussions.

L.I. and U.T. acknowledge funding from the European Research Council (ERC) under the European Union's Horizon 2020 Research and Innovation Program grant agreement No. 101001514. The authors also acknowledge the use of cleanroom facilities at the Binnig and Rohrer Nanotechnology Center (BRNC) at IBM Research -- Zurich and at the FIRST Center for Micro- and Nanoscience at ETH Zurich. A.G. acknowledges AGAUR and FSE+ and Generalitat de Catalunya for financial support under the call FI SDUR 2023 Ref.~CCI 2021ES05FPR011. M.C. and A.D. acknowledge financial support through Grant No. ProyExcel\_00505, funded by Junta de Andaluc\'{\i}a, and Grants No. PID2021-126348NB-I00 and No. PID2024-156257NB-C21, funded by MI-CIU/AEI/10.13039/501100011033 and by ``ERDF/EU''. D.L. acknowledges MCIU/AEI for financial support under Projects No.~PID2022-140407NB-C22 and No.~PID2025-170781NB-C22. I.P. acknowledges support from Ministerio de Ciencia, Innovaci\'on y Universidades MCIU/AEI/FEDER under grant agreement PID2024-156516NB-10, and Generalitat de Catalunya under Program Icrea Acad\`emia and project 2021SGR-673. Development of numerical solvers and simulation code was supported by Anthropic Claude Opus 5.5.  
\end{acknowledgments}
\vspace{0.5em}
\paragraph{Data availability.} The data are available from the authors upon reasonable request.

\bibliographystyle{apsrev4-2}
\bibliography{references.bib}

\clearpage
\onecolumngrid

\renewcommand{\title}[2][]{}
\renewcommand{\author}[2][]{}
\renewcommand{\affiliation}[2][]{}
\renewcommand{\email}[2][]{}
\DeclareRobustCommand{\maketitle}{\begin{center}\textbf{\large Supplemental Material}\end{center}\vspace{1em}}
\renewcommand{\bibliographystyle}[1]{}
\renewcommand{\bibliography}[1]{}

\setcounter{equation}{0}\setcounter{figure}{0}\setcounter{table}{0}\setcounter{section}{0}
\renewcommand{\theequation}{S\arabic{equation}}
\renewcommand{\thefigure}{S\arabic{figure}}
\renewcommand{\thetable}{S\arabic{table}}
\renewcommand{\thesection}{S\arabic{section}}
\renewcommand{\theHequation}{S\arabic{equation}}
\renewcommand{\theHfigure}{S\arabic{figure}}
\renewcommand{\theHtable}{S\arabic{table}}
\renewcommand{\theHsection}{S\arabic{section}}

\title{Supplemental Material for ``Delay-induced current reversal in traveling activity waves''}

\author{Ueli T\"opfer}
    \affiliation{Department of Materials, ETH Zurich, 8049 Zurich, Switzerland}

\author{Adri\`a Garc\'es}
    \affiliation{Departament de F\'\i sica de la Mat\`eria Condensada, Universitat de Barcelona, Mart\'\i{} i Franqu\`es 1, E08028 Barcelona, Spain}
    \affiliation{University of Barcelona Institute of Complex Systems (UBICS), Mart\'\i{} i Franqu\`es 1, E08028 Barcelona, Spain}
    \affiliation{Computing and Understanding Collective Action (CUCA) Lab, Mart\'\i{} i Franqu\`es 1, E08028 Barcelona, Spain}

\author{Miguel Chamorro-Burgos}
    \affiliation{F\'\i sica Te\'orica, Universidad de Sevilla, Apdo.~1065, 41080 Sevilla, Spain}

\author{Demian Levis}
    \affiliation{Departament de F\'\i sica de la Mat\`eria Condensada, Universitat de Barcelona, Mart\'\i{} i Franqu\`es 1, E08028 Barcelona, Spain}
    \affiliation{University of Barcelona Institute of Complex Systems (UBICS), Mart\'\i{} i Franqu\`es 1, E08028 Barcelona, Spain}
    \affiliation{Computing and Understanding Collective Action (CUCA) Lab, Mart\'\i{} i Franqu\`es 1, E08028 Barcelona, Spain}

\author{Alvaro Dom\'\i nguez}
    \affiliation{F\'\i sica Te\'orica, Universidad de Sevilla, Apdo.~1065, 41080 Sevilla, Spain}
    \affiliation{Instituto Carlos I de F\'\i sica Te\'orica y Computacional, 18071 Granada, Spain}

\author{Ignacio Pagonabarraga}
    \affiliation{Departament de F\'\i sica de la Mat\`eria Condensada, Universitat de Barcelona, Mart\'\i{} i Franqu\`es 1, E08028 Barcelona, Spain}
    \affiliation{University of Barcelona Institute of Complex Systems (UBICS), Mart\'\i{} i Franqu\`es 1, E08028 Barcelona, Spain}

\author{Lucio Isa}
    \email{lucio.isa@mat.ethz.ch}
    \affiliation{Department of Materials, ETH Zurich, 8049 Zurich, Switzerland}

\maketitle

\section{Particle Fabrication}
Silica--titania Janus particles were fabricated from monodisperse silica colloids (\SI{2.06}{\micro\metre} diameter, \SI{5}{\percent} w/v, microParticles GmbH, Germany), diluted to \SI{0.5}{\percent} w/v in Milli-Q water. \SI{100}{\micro\litre} of the sonicated suspension were spread onto a glass slide cleaned by \SI{1}{\minute} air-plasma treatment and allowed to dry into a colloidal monolayer. The monolayer was coated with \SI{50}{\nano\metre} of \ce{Ti3O5} by electron-beam evaporation (Plassys MEB700SL) and subsequently annealed at \SI{500}{\degreeCelsius} for \SI{120}{\minute} under \SI{1}{\bar} \ce{O2} in a rapid thermal processor (AS-One 150, ANNEALSYS) to promote full oxidation and conversion to the \ce{TiO2} anatase phase. Particles were retrieved by sonication in Milli-Q water. Prior to each experiment, the dispersion was mixed in equal volumes with a \SI{0.6}{\percent} w/v Pluronic F-127 solution, yielding a final surfactant concentration of \SI{0.3}{\percent} w/v to reduce particle adhesion to the substrate.

\section{Experimental cell}

Observation cells were assembled from two indium--tin oxide coated glass electrodes (\SI{16}{\milli\metre} × \SI{16}{\milli\metre}, \SI{<7}{\ohm\per\square}, präzisions glas \& optik) sandwiched around a \SI{120}{\micro\metre} thick adhesive spacer with a \SI{9}{\milli\metre} circular opening (Grace Bio-Laboratories SecureSeal, USA). The spacer was halved to provide an inlet and outlet, through which \SI{9.5}{\micro\litre} of particle suspension was introduced before sealing with vacuum grease (Krytox GPL205). Electrical contact was made with adhesive copper tape, and an AC electric field was applied via an arbitrary waveform generator (33522A, Agilent) at voltages between \SIrange{1}{10}{\volt} and frequencies between \SIrange{4}{10}{\kilo\hertz}.

\section{Microscopy and illumination}

All experiments were performed on an inverted optical microscope (Eclipse Ti2-E, Nikon) using \SI{20}{\times} (CFI S Plan Fluor ELWD 20XC) and \SI{40}{\times} (CFI S Plan Fluor ELWD 40XC) objectives. UV illumination ($\lambda = \SI{365}{\nano\metre}$, UHP-F-365, Prizmatix) was applied at intensities up to \SI{6.3}{\watt\per\centi\metre\squared} and spatially patterned within the field of view using a DMD illuminator (Polygon 1000, Mightex, USA). Since the DMD modulates light through binary (on/off) micromirrors, grayscale intensity patterns were generated by spatial dithering over $9\times9$ mirror supercells, allowing the traveling wave's continuous intensity profile to be approximated at the resolution of one supercell. A DAPI filter cube suppressed reflected UV light to prevent camera saturation. Image sequences ($2160 \times 2560$ pixels) were acquired with an sCMOS camera (Andor Zyla) at \SI{10}{\per\second}.

\section{Notation}

Position in the lab frame is denoted by $x$. The wave-comoving phase is
\begin{equation}
\zeta \equiv \mathrm{mod}\!\left(\frac{x - x_{\mathrm{wave}}(t)}{\ell} + \frac14,\,1\right) \in [0,1),
\end{equation}
where $x_{\mathrm{wave}}(t)$ is the position of the traveling light pattern and $\ell$ the wavelength; the offset $1/4$ is chosen so that the wave intensity peaks at $\zeta=0.5$ rather than at $\zeta=0.25$, its location in the raw lab-frame coordinate. In this convention the driving profile is
\begin{equation}
u(\zeta) = \tfrac12\big(1-\cos 2\pi\zeta\big),
\end{equation}
zero at $\zeta=0$ and $\zeta=1$, peaking at $\zeta=0.5$. These conventions are used throughout both this Supplemental Material and the main text.

\section{Particle tracking and analysis}
 
Particle positions were detected in each frame and linked into trajectories using custom MATLAB routines. The self-propulsion speed $v_0$ was determined by fitting the particle ensemble's MSD, computed from \SI{20}{\second} of full-field UV illumination prior to pattern onset, to the ABP model
\begin{equation}
\text{MSD}(\Delta t) = \left(4D + 2v_0^2 D_r^{-1}\right)\Delta t + 2v_0^2 D_r^{-2}\!\left(e^{-\Delta t D_r} - 1\right),
\end{equation}
with $D$ and $D_r$ as free parameters, giving $D = \qty{0.11}{\um^2\per\second}$ and $D_r = 1/\tau_r = \qty{0.16}{\per\second}$ ($\tau_r = \qty{6.3}{\second}$), consistent with the values reported in Ref.~\cite{Tpfer2025}.\\

Instantaneous velocities were extracted as a function of wave phase using a fixed-lag, phase-binned estimator. For each particle track, the velocity was computed over a fixed window of $n_{\mathrm{step}}=3$ frames rather than frame-to-frame, short enough that the particle does not appreciably sweep across the wave during the measurement, but long enough to average out pixel-level tracking noise. Each sample was tagged with the wave-comoving phase $\zeta$, evaluated at the window's starting position and time using the logged position of the light pattern. All $(\zeta, v_x, v_y)$ samples from every particle and every valid window across a recording were pooled into a single set, from which the speed $|v| = \sqrt{v_x^2+v_y^2}$ was also computed per sample. The pooled samples were then binned by $\zeta$ into $30$ equal-width bins spanning one full period, and the mean and standard error within each bin yield the phase-resolved velocity profiles $v_\zeta$, $v_y(\zeta)$, and $|v|(\zeta)$ reported throughout the main text. No drift correction was applied to these phase-resolved profiles.\\
 
Net drift velocities are only a few hundred \si{\nm\per\second}, so even a small background motion of the sample or stage can bias them. We therefore subtract this background motion before computing the net drift velocity. Because it need not be uniform across the field of view, we estimate it locally. We divide the field of view into a $2\times2$ grid of boxes, with neighboring boxes overlapping by 50\%. Within each box, we collect the velocities of all particles whose displacement starts inside it, computed over lag times from \SIrange{1}{11}{\second} and pooled together. The velocity distribution in each box is dominated by a peak from particles that are not actively transported, and this peak is offset from zero only by the background motion. We locate it as the maximum of a Gaussian kernel density estimate closest to zero velocity, and take its position as the box's background velocity. Subtracting this background velocity from all velocities in the box and averaging gives the box's corrected net drift velocity. The net drift velocity and its uncertainty are the mean and standard deviation of this quantity across all boxes.

\section{Simulation model}
 
Particle dynamics were modeled using overdamped active Brownian particles (ABPs) in two dimensions. Each particle $i$ ($i=1,\ldots,N$) is described by its position $\mathbf{r}_i$ and orientation angle $\theta_i$, evolving according to
\begin{align}
\dot{\mathbf{r}}_i &= v_{\mathrm{a},i}(t)\,\hat{\mathbf{p}}_i + \sqrt{2D}\;\boldsymbol{\eta}_i^{(t)}, \\
\dot{\theta}_i &= \sqrt{2D_r}\;\eta_i^{(r)},
\end{align}
where $v_{\mathrm{a},i}$ is the particle's active (self-propulsion) speed, $\hat{\mathbf{p}}_i = (\cos\theta_i,\sin\theta_i)$ is the propulsion direction, $D$ and $D_r$ are the translational and rotational diffusion coefficients, and the noise terms $\boldsymbol{\eta}_i^{(t)}$, $\eta_i^{(r)}$ are independent Gaussian white-noise processes with zero mean and unit-time correlator; free rotational diffusion gives each particle a persistence time $\tau_r = 1/D_r$. Periodic boundary conditions were applied in both directions.
 
The traveling activity wave sets a local target speed
\begin{equation}
u_{\mathrm{target},i}(t) = \tfrac12 u_{\max}\!\left[1 + \sin\!\left(\frac{2\pi(x_i - c\,t)}{\ell}\right)\right],
\end{equation}
with wave speed $c$ and wavelength $\ell$. Rather than adapting to $u_{\mathrm{target}}$ instantaneously, each particle's active speed relaxes toward it with a first-order lag,
\begin{equation}
\frac{dv_{\mathrm{a},i}}{dt} = \frac{u_{\mathrm{target},i}(t) - v_{\mathrm{a},i}}{\tau_i}, \qquad
\tau_i = \begin{cases} \tau_{\mathrm{on}} & \text{if } u_{\mathrm{target},i}(t) \geq v_{\mathrm{a},i}(t), \\ \tau_{\mathrm{off}} & \text{if } u_{\mathrm{target},i}(t) < v_{\mathrm{a},i}(t), \end{cases}
\end{equation}
integrated exactly over each time step. All simulations reported here used $\tau_{\mathrm{on}} = 0$ (instantaneous rise), with $\tau_{\mathrm{off}}\equiv\tau$ the free delay parameter carrying the entire asymmetric response.
 
The maximum active speed $u_{\max}$ (\SIrange{0}{10}{\um\per\second}), wave speed $c$ (\SIrange{0}{20}{\um\per\second}), and delay $\tau$ (\SIrange{0}{8}{\second}) were swept at fixed wavelength $\ell = \qty{200}{\um}$. Simulations were performed in a periodic square domain of side $L = \qty{800}{\um}$ containing $N = 640$ particles (number density $N/L^2 = \num{0.001}\,\si{\per\um^2}$). Translational and rotational diffusion coefficients were fixed via the Stokes--Einstein relation for a particle radius $r = \qty{1.03}{\um}$: $D = \qty{0.11}{\um^2\per\second}$ and $D_r = \qty{0.16}{\per\second}$ (persistence time $\tau_r = 1/D_r \approx \qty{6.25}{\second}$), using $k_BT = \qty{0.00409}{\pico\newton\um}$ and mobility $\mu = \qty{26.9}{\um\per\pico\newton\per\second}$.
 
Equations of motion were integrated using the Euler--Maruyama scheme with a time step $\Delta t = \qty{1}{ms}$, chosen to satisfy $\Delta t \ll \tau_r$ and $\Delta t \ll \ell/c$. Each simulation ran for $10^6$ steps (total simulated time $T = \qty{1000}{\second}$), with configurations saved every 100 steps ($10^4$ saved frames). Simulated trajectories were then converted to the wave-comoving phase $\zeta$ and phase-binned identically to the experimental tracking data (see Particle tracking and analysis, above), to extract the local mean velocity $v_\zeta$, density, and polarization.

\section{Propulsion response to illumination switching}
 
\begin{figure}[ht]
\centering
\includegraphics[width=0.4\columnwidth]{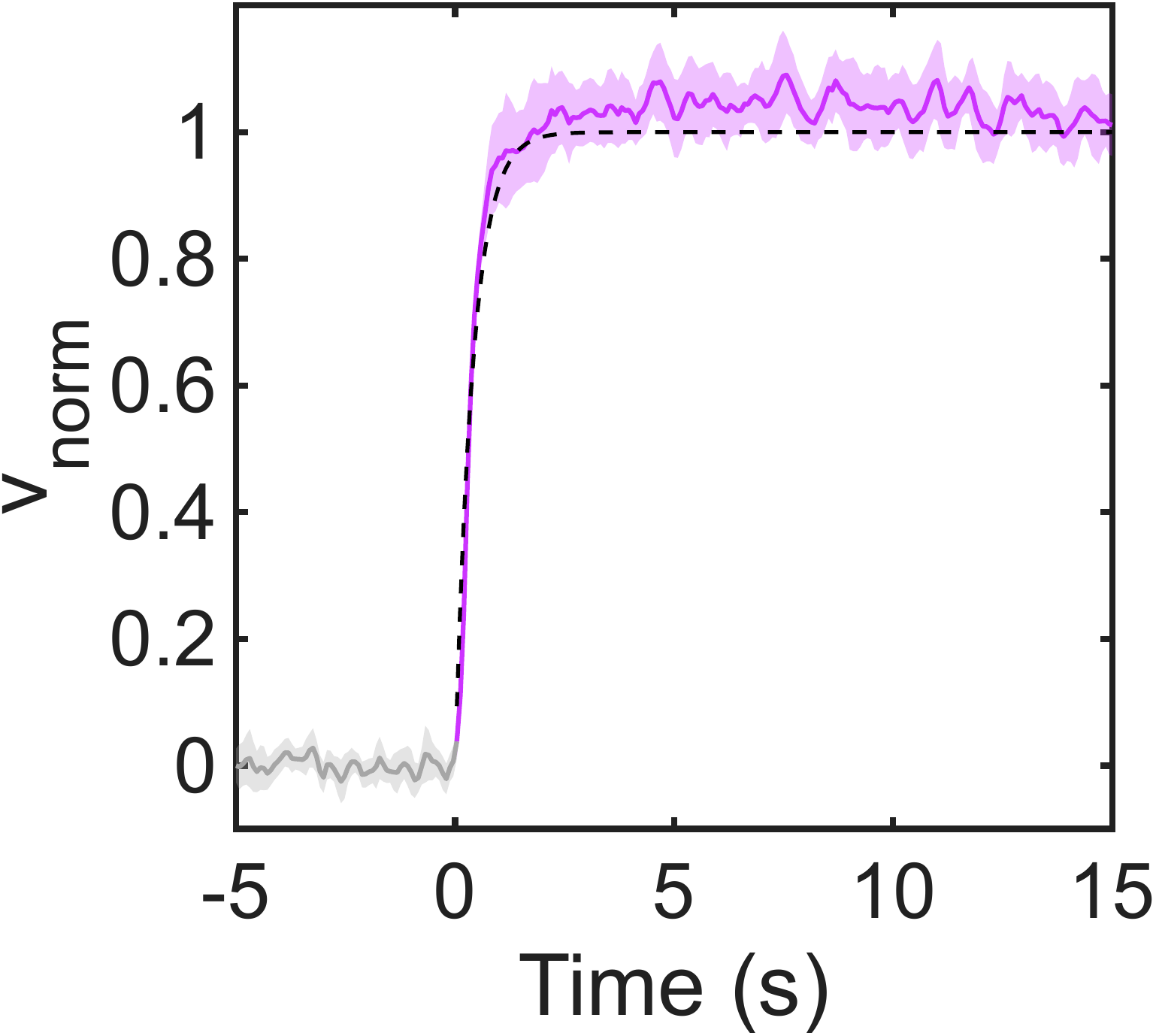}
\caption{\textbf{Switch-on response of the active speed.}
Active speed following a spatially uniform switch-on of UV illumination, normalized and averaged over three illumination cycles across four experimental cells; shaded regions are 95\% confidence intervals. The rise reaches steady state within \SI{0.4}{\second}, close to the tracking's temporal resolution.}
\label{fig:switch_on}
\end{figure}

The active speed's response to a step change in UV illumination was characterized separately from the traveling-wave experiments, under spatially uniform illumination. Switching illumination on from a dark baseline, the active speed rises to its steady-state value within \SI{0.4}{\second}, close to the \SI{0.3}{\second} window used for velocity extraction, so this rise cannot be resolved as a genuine exponential relaxation and is only bounded from above by this timescale. This transient was normalized and averaged over three illumination cycles across four cells, with shaded regions in Fig.~\ref{fig:switch_on} indicating 95\% confidence intervals.
 
\section{Driving profile measurement}

The target-speed profile imposed by the traveling wave is not assumed to be sinusoidal, and is instead measured from the particles' own phase-resolved speed. As the wave passes over a particle, the particle first experiences increasing intensity up to the wave peak, and then decreasing intensity after it. We measure the driving profile only from the first, rising part, using the window $\zeta\in[0.45,0.95]$. On this half, where the active speed rises with the increasing intensity, the effect of the delay is minimized (Propulsion response to illumination switching, above), so the rising-half speed closely approximates $u_{\mathrm{target}}(\zeta)$. On the falling half, in contrast, the speed lags the target by the delay $\tau$ and therefore reflects both the driving shape and the delay. A further advantage of extracting the driving profile from the particles' speed rather than from the light field is that it automatically captures the profile the particles actually respond to, including any distortion from the DMD dithering and any nonlinearity in the speed--intensity response.\\

\paragraph{Correction for Rice-distributed speed measurements.} The measured speed $|v|=\mathrm{hypot}(v_x^{\mathrm{meas}},v_y^{\mathrm{meas}})$ is reconstructed from finite-difference position increments, with each Cartesian velocity component subject to independent, approximately Gaussian measurement noise of standard deviation $\sigma = \sqrt{2D/\Delta t_{\mathrm{lag}}}$, where $D$ is the translational diffusion coefficient and $\Delta t_{\mathrm{lag}}$ the lag time used for the finite difference. If $v_x^{\mathrm{meas}}\sim\mathcal{N}(\nu\cos\theta,\sigma^2)$ and $v_y^{\mathrm{meas}}\sim\mathcal{N}(\nu\sin\theta,\sigma^2)$ for a true active speed $\nu$, the magnitude $|v|$ follows a Rice distribution $\mathrm{Rice}(\nu,\sigma)$, whose mean is systematically biased above $\nu$:
\begin{equation}
\mathbb{E}[|v|] = \sigma\sqrt{\tfrac{\pi}{2}}\;e^{-s}\Big[(1+2s)\,I_0(s) + 2s\,I_1(s)\Big], \qquad s \equiv \frac{\nu^2}{4\sigma^2},
\label{eq:rice}
\end{equation}
with $I_0,I_1$ modified Bessel functions of the first kind.\\

\begin{figure}[ht]
\centering
\includegraphics[width=0.6\textwidth]{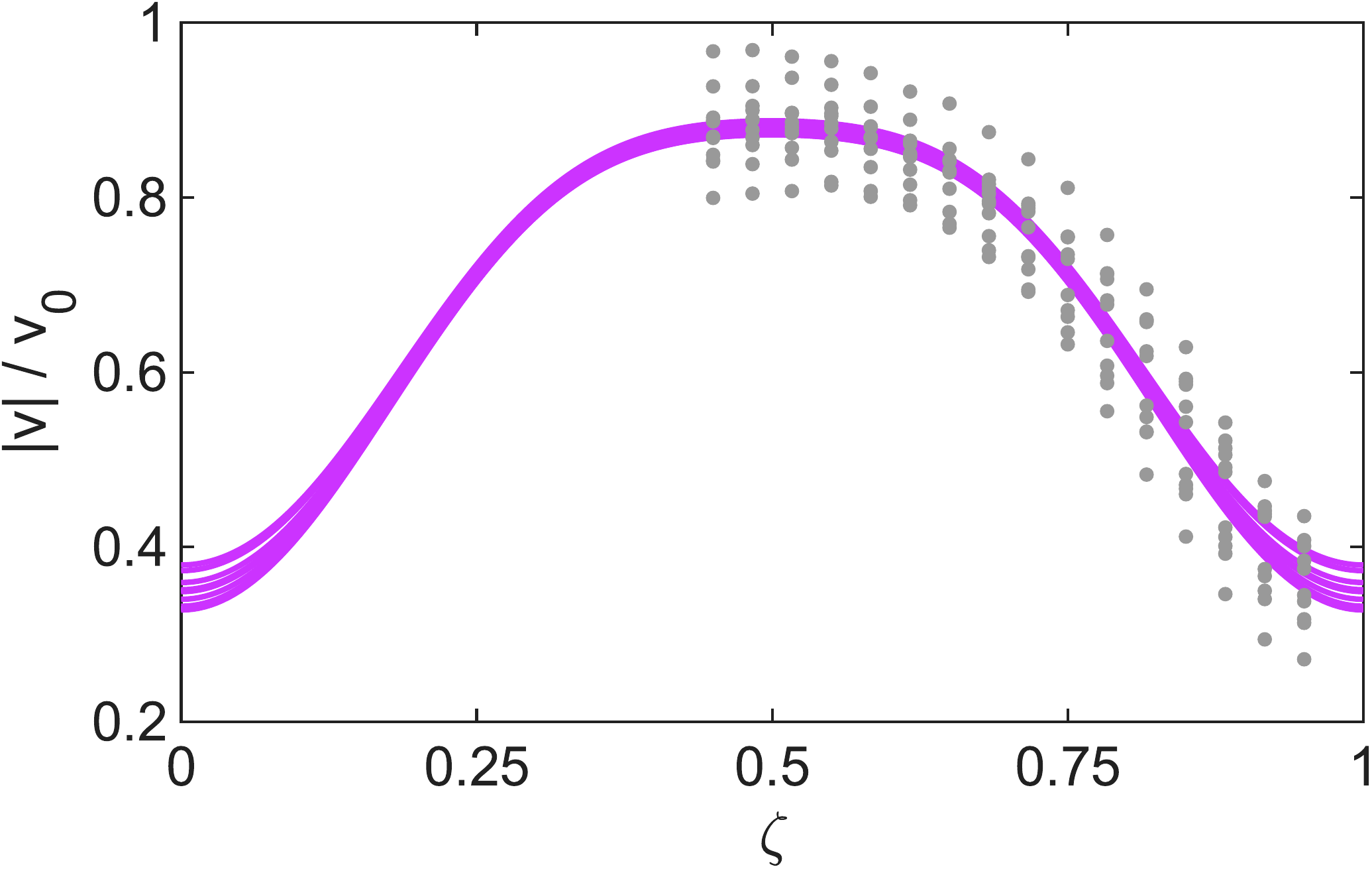}
\caption{\textbf{Driving profile from the rising half.}
Gray: mean speed $|v|/v_0$ over the fit window $\zeta\in[0.45,0.95]$, for every recording used in the fit, each normalized by its own pre-pattern active speed $v_0$. Purple: the shared driving profile $v_0 f(\zeta)$, first forward-biased through Eq.~(\ref{eq:rice}) with each recording's noise level and then normalized by $v_0$, one line per recording. The collapse of recordings with different $c$ and $v_0$ onto one curve supports a single driving shape whose amplitude is set by $v_0$ alone.}
\label{fig:driving_shape}
\end{figure}

\paragraph{Shared driving profile.} We pool the rising-half, phase-binned mean speed from 11 recordings spanning $c=\SIrange{4.1}{7.7}{\um\per\second}$ and $v_0=\SIrange{4.29}{7.59}{\um\per\second}$, each normalized by its own $v_0$, the active speed obtained from the MSD fit under full-field illumination before pattern onset (Particle tracking and analysis, above). We describe all recordings with one shared dimensionless shape $f(\zeta)$, so that $u_{\mathrm{target}}(\zeta)=v_0\,f(\zeta)$, and represent $f$ by a second-order Fourier series. Its coefficients are determined by a least-squares fit of the forward-biased model speed, i.e.\ $v_0 f(\zeta)$ passed through Eq.~(\ref{eq:rice}), to the measured mean speed in each phase bin. The resulting driving profile, Fig.~\ref{fig:driving_shape}, is used for all theory--experiment comparisons (Fig.~\ref{fig:phase_curve_parity}a).
 
\section{Model}
 
\paragraph{Moment expansion.} The particle's Langevin dynamics in the wave-comoving frame $z\equiv x-ct$, and the resulting Fokker--Planck equation for the joint probability density $\psi(z,\theta,v,t)$,
\begin{equation}
\frac{\partial \psi}{\partial t} = -\frac{\partial}{\partial z}\Big[(v\cos\theta - c)\,\psi\Big] + D\,\frac{\partial^2 \psi}{\partial z^2} + D_r\,\frac{\partial^2 \psi}{\partial \theta^2} + \frac{\partial}{\partial v}\left[\left(\frac{v-u(z)}{\tau(z,v)}\right)\psi\right], \label{eq:fp}
\end{equation}
are given in the main text. We reduce Eq.~(\ref{eq:fp}) by angular-moment truncation to a small, closed linear system, solved directly as a single sparse linear algebra problem in the wave-comoving frame. $\psi$ is expanded in harmonics of $\theta$. Since the wave profile is translation-invariant along the y axis, $\psi$ contains no $\sin(k\theta)$ terms, and the expansion is a pure cosine series,
\begin{equation}
\psi(z,\theta,v,t) = \frac{1}{2\pi}\left[\Psi_0(z,v,t) + 2\sum_{k=1}^{\infty}\Psi_k(z,v,t)\,\cos(k\theta)\right],
\end{equation}
with coefficients
\begin{equation}
\Psi_0(z,v) = \int \psi\,d\theta, \qquad
\Psi_1(z,v) = \int \cos\theta\,\psi\,d\theta, \qquad
\Psi_2(z,v) = \int \cos 2\theta\,\psi\,d\theta,
\end{equation}
the density, and the polar and nematic order parameters of the heading distribution, respectively: $\Psi_1$ is nonzero under a genuine head-tail asymmetry, with more particles oriented one way along an axis than in the opposite direction, while $\Psi_2$, familiar from liquid-crystal order, is nonzero under alignment onto an axis without distinguishing its two directions, and can be nonzero even when $\Psi_1=0$. No explicit alignment interaction appears in Eq.~(\ref{eq:fp}); persistent orientation enters only through advection and isotropic rotational diffusion. We look for stationary states in the wave-comoving frame, so we set $\left.\partial_t\right|_z\to 0$ throughout. Because $\{1,\cos\theta,\cos2\theta,\dots\}$ are mutually orthogonal on $[0,2\pi)$, projecting Eq.~(\ref{eq:fp}) onto $\cos(k\theta)$ recovers the equation obeyed by $\Psi_k$ directly from the series above, rather than a separate calculation: multiplying Eq.~(\ref{eq:fp}) by $\cos(k\theta)$ for $k=0,1,2$ and integrating over $\theta$, using $\cos\theta\cos(k\theta) = \tfrac12[\cos((k-1)\theta)+\cos((k+1)\theta)]$ and $\int\cos(k\theta)\,\partial_\theta^2\psi\,d\theta = -k^2\int\cos(k\theta)\,\psi\,d\theta$, and closing the hierarchy by setting $\Psi_3\equiv0$, gives
\begin{align}
-\partial_z\Big[v\Psi_1 - c\Psi_0\Big] + D\partial_z^2\Psi_0 + \partial_v\left[\frac{v-u(z)}{\tau(z,v)}\Psi_0\right] &= 0, \label{eq:Psi0}\\
-\partial_z\Big[\tfrac12 v(\Psi_0+\Psi_2) - c\Psi_1\Big] + D\partial_z^2\Psi_1 - D_r\Psi_1 + \partial_v\left[\frac{v-u(z)}{\tau(z,v)}\Psi_1\right] &= 0, \label{eq:Psi1}\\
-\partial_z\Big[\tfrac12 v\Psi_1 - c\Psi_2\Big] + D\partial_z^2\Psi_2 - 4D_r\Psi_2 + \partial_v\left[\frac{v-u(z)}{\tau(z,v)}\Psi_2\right] &= 0. \label{eq:Psi2}
\end{align}
 
\begin{figure}[ht]
\centering
\includegraphics[width=0.95\columnwidth]{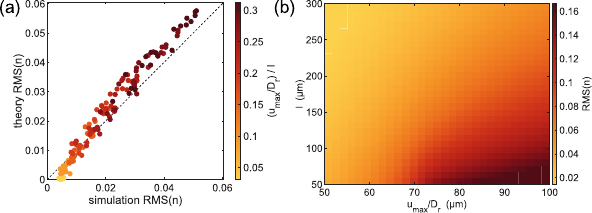}
\caption{\textbf{Nematic order requires closing the hierarchy at $\Psi_3\equiv0$.}
(\textbf{a}) Root-mean-square (RMS) nematic order $\langle\cos2\theta\rangle(\zeta)$ measured directly in simulated trajectories (horizontal axis) compared against the $\Psi_3\equiv0$ model's own prediction (vertical axis; the $\Psi_2\equiv0$ closure predicts this identically zero and so has no prediction to show here), for every tested combination of $u_{\max}$, $\tau\in\{0,2,4,8\}\,\si{\second}$, and $c\in\{2,4,6,8,10\}\,\si{\um\per\second}$, colored by the persistence-length ratio $\ell_p/\ell\equiv u_{\max}/(D_r\ell)$. Dashed line: $y=x$. Agreement confirms the model correctly predicts the amplitude of the real nematic signal. (\textbf{b}) Root-mean-square nematic order predicted by the $\Psi_3\equiv0$ model (color), swept over persistence length $\ell_p\equiv u_{\max}/D_r$ and wavelength $\ell$ independently, at fixed particle radius $R=\SI{1.03}{\um}$ (fixing $D_r$) and wave speed $c=\SI{10}{\um\per\second}$, in the zero-delay limit.}
\label{fig:rms_p2}
\end{figure}

\paragraph{Why the nematic order matters.} Each equation is sourced by a spatial derivative of the moment one order below: $\Psi_1$'s equation is driven by $\partial_z[v\Psi_0]$, and $\Psi_2$'s in turn by $\partial_z[v\Psi_1]$, each of order $u_{\max}\Psi_0/\ell$ and $u_{\max}\Psi_1/\ell$ respectively, since $\psi$ varies over the wavelength $\ell$. Balancing this forcing against the dominant relaxation term at each order, $D_r\Psi_1$ and $4D_r\Psi_2$, gives the rough scaling $\Psi_1/\Psi_0 \sim \Psi_2/\Psi_1 \sim u_{\max}/(\ell D_r) = \ell_p/\ell$, so $\Psi_2/\Psi_0 \sim (\ell_p/\ell)^2$, with $\ell_p\equiv u_{\max}/D_r$ the persistence length. The cascade is therefore weak, and truncating at low order safe, when $\ell_p\ll\ell$, but strong once $\ell_p\gtrsim\ell$, matching the same intermediate-scale regime in which higher-order closures become necessary for plain active Brownian particles~\cite{Gautry2026}. This delay-aware, steady-state $(z,v)$ system, closed at $\Psi_3\equiv0$, is the model used throughout. Closing instead at $\Psi_2\equiv0$ predicts the nematic order to be exactly zero, independent of any residual-based accuracy comparison; we test this directly against real simulated trajectories in Fig.~\ref{fig:rms_p2}(a), which indeed shows the RMS nematic order growing with increasing $\ell_p/\ell$ (colorbar), and confirms that closing the hierarchy at $\Psi_3\equiv0$ reproduces this simulated RMS nematic order to a good agreement. We map the predicted size of the effect directly across the $(\ell_p,\ell)$ plane in Fig.~\ref{fig:rms_p2}(b), where the RMS nematic order is largest for large $\ell_p$ and small $\ell$, consistent with the scaling $\Psi_2/\Psi_0\sim(\ell_p/\ell)^2$ derived above.
 
\paragraph{Zero-delay limit.} In this limit, the active speed is adiabatically eliminated to the local target, $v=u(z)$, and $v$ is no longer an independent coordinate: substituting it directly into Eq.~(\ref{eq:fp}) removes the $\partial_v[\dots]$ term entirely, and repeating the same angular-moment projection on this simplified equation gives a closed system of ordinary differential equations in $z$ alone, obtained from Eqs.~(\ref{eq:Psi0})--(\ref{eq:Psi2}) by dropping the delay term and replacing $v$ with $u(z)$ throughout.
 
\paragraph{Delay-aware case.} When $\tau>0$, the active speed can no longer be eliminated as $v=u(z)$: it remains a genuine internal coordinate, and Eqs.~(\ref{eq:Psi0})--(\ref{eq:Psi2}) retain their full $v$-dependence through $\tau(z,v)$, which switches between $\tau_{\mathrm{on}}$ and $\tau_{\mathrm{off}}$ depending on whether the local target speed is rising or falling relative to $v$. This makes the delay-aware steady-state system a genuine two-dimensional problem in $(z,v)$, not reducible back to the one-dimensional system above.
 
\paragraph{Discretization.} Both cases are linear, since the moments couple only linearly, so after discretization the entire problem reduces to a sparse linear system $A\mathbf{u}=\mathbf{b}$. The $z$-derivatives (diffusion and advection) are approximated with central differences on a periodic grid over one wavelength, $z\in[0,\ell)$, in both cases. The additional $v$-advection term in the delay-aware case is discretized instead with an upwind finite-volume scheme, since that direction is not periodic and has physical boundaries at $v=0$ and $v=u_{\max}$; central differencing is unstable for a purely advective term, whereas upwinding (drawing the flux from whichever neighboring cell the drift originates from) is stable. In both cases, the resulting matrix $A$ is singular by construction, since the moment equations fix only the shape of the distribution and not its overall scale; one row of $A$ is therefore overwritten by the normalization constraint ($\int_0^\ell \Psi_0\,dz=1$ in the zero-delay case, $\int_0^\ell\!\!\int_0^{u_{\max}} \Psi_0\,dv\,dz=1$ in the delay-aware case), with the corresponding entry of $\mathbf{b}$ set to $1$. The system is then solved directly by sparse LU decomposition, with no iteration or time-stepping required.
 
\paragraph{Observables.} Once $\Psi_0,\Psi_1,\Psi_2$ are known, all reported quantities follow by quadrature over $z$: the density $\Psi_0(z)$, the polar order $\Psi_1(z)$, the flux $q(z)=\int v\,\Psi_1\,dv$ (or $q(z)=v(z)\,\Psi_1(z)$ in the zero-delay case, where no separate $v$-integral is needed), and the local mean velocity $v_z(z)=q(z)/\Psi_0(z)$. For comparison with experiment and simulation we report these on the dimensionless, periodic phase $\zeta\equiv z/\ell$ (Notation, above) used throughout the main text. The density picks up a Jacobian factor to remain normalized, $\rho(\zeta)\equiv\ell\,\Psi_0(z=\ell\zeta)$, so that $\int_0^1\rho(\zeta)\,d\zeta=1$; the mean velocity, an intensive ratio, is unaffected by this rescaling, $v_\zeta\equiv v_z(z=\ell\zeta)$. The net drift velocity then follows either as $v_{\mathrm{transport}}=\int_0^1\rho(\zeta)\,v_\zeta\,d\zeta$, matching the main text since $\rho v_\zeta=q(\ell\zeta)$, or equivalently, without first forming $\rho(\zeta)$, directly as $v_{\mathrm{transport}}=\int_0^\ell q(z)\,dz$. Because the system is solved exactly (given $D$, $D_r$, $c$, $\ell$, $u_{\max}$, and $\tau$) rather than sampled stochastically, this constitutes a closed-form numerical solution of the truncated-moment system rather than a simulation.

\section{Model validation}

\begin{figure}[ht]
\centering
\includegraphics[width=0.9\columnwidth]{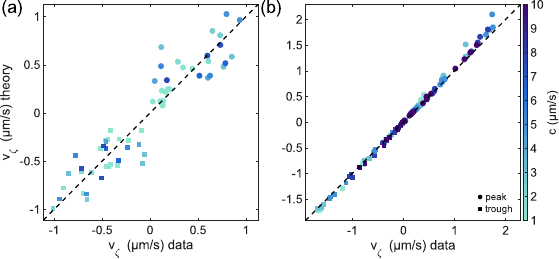}
\caption{\textbf{Theory--experiment and theory--simulation parity for the local mean velocity.}
Predicted (theory) vs. measured or simulated peak and trough of $v_\zeta(\zeta)$, one peak and one trough per run; both are extracted by fitting a periodic smoothing spline to the phase-binned curve and reading off its extrema, applied identically to the theory's noise-free solution for consistency. Circles: peak; squares: trough; color: wave speed $c$, \SIrange{1}{10}{\um\per\second}. Dashed line: $y=x$.
(\textbf{a}) Experiment, all runs, using each run's own measured wave speed and particle speed, fixed $D$ and $D_r$, fixed delay $\tau=\SI{2}{\second}$, and the empirically measured driving profile (Driving profile measurement, above).
(\textbf{b}) Simulation, a representative subset spanning $u_{\max}\in\{1,3,5,8\}\,\si{\um\per\second}$, $\tau\in\{0,2,4,8\}\,\si{\second}$, and $c\in\{1,3,5,8,10\}\,\si{\um\per\second}$.}
\label{fig:phase_curve_parity}
\end{figure}

Each run is reduced to the peak and trough of its $v_\zeta(\zeta)$ curve, and compared against the theory evaluated for that same run (Fig.~\ref{fig:phase_curve_parity}).

\pagebreak

\bibliographystyle{apsrev4-2}
\bibliography{references.bib}

\end{document}